\documentclass[%
reprint,
superscriptaddress,
showpacs,
showkeys,
 amsmath,amssymb,
 aps,
prc,
floatfix,
altaffilletter,
letter
]{revtex4-1}

\usepackage{graphicx}
\usepackage{dcolumn}
\usepackage{bm}
\usepackage{multirow}
\usepackage{amssymb}
\usepackage{mathrsfs}

\begin{document}

\title{$\Sigma^- p$ emission rates in $K^-$ absorptions at rest on  
$^6$Li, $^7$Li, $^{9}$Be, $^{13}$C and $^{16}$O}

\newcommand*{\UNIBA}{Dipartimento di Fisica, Universit\`a di Bari, 70125
Bari, Italy}
\newcommand*{\UNIBAindex}{1}
\affiliation{\UNIBA}
\newcommand*{\INFNBA}{I.N.F.N. Sezione di Bari, 70125 Bari, Italy}
\newcommand*{\INFNBAindex}{2}
\affiliation{\INFNBA}
\newcommand*{\UNIBS}{Dipartimento di Ingegneria Meccanica e Industriale, 
25123  Brescia, Italy}
\newcommand*{\UNIBSindex}{3}
\affiliation{\UNIBS}
\newcommand*{\INFNLNF}{I.N.F.N. Laboratori Nazionali di Frascati, 00044
Frascati, Italy}
\newcommand*{\INFNLNFindex}{4}
\affiliation{\INFNLNF}
\newcommand*{\UNIKYO}{Department of Physics, Kyoto University, Kitashirakawa, 
Kyoto 606-8502 Japan}
\newcommand*{\UNIKYOindex}{5}
\affiliation{\UNIKYO}
\newcommand*{\INFNPV}{I.N.F.N. Sezione di Pavia, 27100 Pavia, Italy}
\newcommand*{\INFNPVindex}{6}
\affiliation{\INFNPV}
\newcommand*{\RIKEN}{RIKEN, Wako, Saitama 351-0198, Japan}
\newcommand*{\RIKENindex}{7}
\affiliation{\RIKEN}
\newcommand*{\POLITO}{Dipartimento di Fisica, Politecnico di Torino, 10129 Torino, Italy}
\newcommand*{\POLITOindex}{8}
\affiliation{\POLITO}
\newcommand*{\UNITO}{Dipartimento di Fisica, Universit\`a di Torino, 10125 
Torino, Italy}
\newcommand*{\UNITOindex}{9}
\affiliation{\UNITO}
\newcommand*{\INAF}{I.N.A.F.-I.F.S.I., Sezione di Torino, 10133 Torino, Italy}
\newcommand*{\INAFindex}{10}
\affiliation{\INAF}
\newcommand*{\INFNTO}{I.N.F.N. Sezione di Torino, 10125 Torino, Italy}
\newcommand*{\INFNTOindex}{11}
\affiliation{\INFNTO}
\newcommand*{\UNITS}{Dipartimento di Fisica, Universit\`a di Trieste, 34127
Trieste, Italy}
\newcommand*{\UNITSindex}{12}
\affiliation{\UNITS}
\newcommand*{\INFNTS}{I.N.F.N. Sezione di Trieste, 34127 Trieste, Italy}
\newcommand*{\INFNTSindex}{13}
\affiliation{\INFNTS}


\newcommand*{\corr}{Corresponding author. \\
E-mail: filippi@to.infn.it; Fax: +39.011.6707324.}
\newcommand*{\dead}{Deceased}
\newcommand*{\CEA}{Current address: CEA/SACLAY, DSM/Irfu/SACM, 91191 Gif-sur-Yvette, France}
\newcommand*{\brasil}{Current address: Universidade de S\~ao Paulo, 05508-070 S\~ao Paulo, Brazil}

\author{M.~Agnello}\affiliation{\POLITO}\affiliation{\INFNTO}
\author{L.~Benussi}\affiliation{\INFNLNF}
\author{M.~Bertani}\affiliation{\INFNLNF} 
\author{G.~Bonomi}\affiliation{\UNIBS}\affiliation{\INFNPV} 
\author{E.~Botta}\affiliation{\UNITO}\affiliation{\INFNTO}
\author{M.~Bregant}\altaffiliation{\brasil}\affiliation{\INFNTS}
\author{T.~Bressani}\affiliation{\UNITO}\affiliation{\INFNTO}
\author{S.~Bufalino}\affiliation{\INFNTO}
\author{L.~Busso}\affiliation{\UNITO}\affiliation{\INFNTO}
\author{D.~Calvo}\affiliation{\INFNTO}
\author{P.~Camerini}\affiliation{\UNITS}\affiliation{\INFNTS}
\author{B.~Dalena}\altaffiliation{\CEA}\affiliation{\UNIBA}\affiliation{\INFNBA}
\author{F.~De~Mori}\affiliation{\UNITO}\affiliation{\INFNTO}
\author{G.~D'Erasmo}\affiliation{\UNIBA}\affiliation{\INFNBA}
\author{A.~Feliciello}\affiliation{\INFNTO}
\author{A.~Filippi}\altaffiliation{\corr}\affiliation{\INFNTO}
\author{E.M.~Fiore}\affiliation{\UNIBA}\affiliation{\INFNBA} 
\author{A.~Fontana}\affiliation{\INFNPV} 
\author{H.~Fujioka}\affiliation{\UNIKYO}
\author{P.~Genova}\affiliation{\INFNPV} 
\author{P.~Gianotti}\affiliation{\INFNLNF} 
\author{N.~Grion}\affiliation{\INFNTS}
\author{V.~Lucherini}\affiliation{\INFNLNF}
\author{S.~Marcello}\affiliation{\UNITO}\affiliation{\INFNTO} 
\author{O.~Morra}\affiliation{\INAF}
\author{T.~Nagae}\affiliation{\UNIKYO} 
\author{H.~Outa}\affiliation{\RIKEN}
\author{A.~Pantaleo}\altaffiliation{\dead}\affiliation{\INFNBA}
\author{V.~Paticchio}\affiliation{\INFNBA}
\author{S.~Piano}\affiliation{\INFNTS}
\author{R.~Rui}\affiliation{\UNITS}\affiliation{\INFNTS}
\author{G.~Simonetti}\affiliation{\INFNBA}
\author{R.~Wheadon}\affiliation{\INFNTO} 
\author{A.~Zenoni}\affiliation{\UNIBS}\affiliation{\INFNPV}

\collaboration{FINUDA Collaboration}\noaffiliation

\date{\today}


\begin{abstract}
An experimental 
study of the $K^-_{stop}A\rightarrow \Sigma^- p A'$ reaction on 
$A=^6$Li, $^7$Li, $^9$Be, $^{13}$C and $^{16}$O $p$-shell nuclei is
presented. The data were collected by the FINUDA spectrometer operating
at the DA$\Phi$NE $\phi$-factory (LNF-INFN, Italy).  
Emission rates for the reaction in the mentioned nuclei are measured and
compared with the few existing data. The spectra of
several observables are discussed; indications of Quasi-Free 
absorptions by a $(np)$ pair embedded in the $A$ nucleus
can be obtained from the study of the missing mass distributions. 
\end{abstract}


\maketitle


\hyphenation{FI-NU-DA}

\section{\label{sec:intro}Introduction}

The existing measurements of kaon absorption reactions by two or few
nucleons are scarce and, for the largest part, dated. Only one paper describes
in good detail the interaction of $K^-$'s at rest on $^4$He  
in bubble chamber \cite{re:katz}. More recently, data have been collected
on $^4$He by the E549 experiment at KEK \cite{re:e549}.
Just few other data, with large or even
no quoted errors, exist. They were taken in emulsion 
experiments, which 
studied the interaction of $K^-$'s with nuclei heavier than $^{16}$O \cite{re:emulsio},
or in bubble chambers filled with hydrocarbon mixtures or Neon \cite{re:bubble}.

The existing measurements suggest a surface behavior of 
the kaonic absorption; they indicate that the contribution of kaonic
absorption reactions  containing at least one nucleon in the final state 
is as sizeable as $\sim 20$\%, 
in medium-heavy nuclei \cite{re:katz, re:bubble}.
Given such a high rate, a more detailed knowledge on the absorption
features is desirable; in fact, 
these processes represent the main background
under the signals of hypernuclei formation in kaon induced
reactions \cite{re:germanoplb}.
Hypernuclear capture rates are, in comparison, at 
least one order of magnitude smaller. 
Therefore, reliable studies of hypernuclei formed by kaons interacting at rest 
with rates of the
order of $10^{-4}-10^{-5}/K^{-}_{stop}$, or of  
decays of hypernuclei in rare modes \cite{re:HypX},
need a precise definition of such background reactions.

More detailed information on the rates of kaon induced reactions in nuclei
is also relevant for astrophysical investigations, for instance concerning 
the composition of possible 
compact stars with hyperonic content, an issue related to the so-called
``hyperon puzzle'' \cite{re:hypstars}.
The presence of hyperons in dense nuclear matter 
could have sizeable effects
on the softness of the compact star Equation of State (EoS); moreover, 
their many-body interactions with nucleons could have a direct impact 
on the possible maximum value of the star mass. The models 
based on standard nuclear physics approaches
rely heavily on the still approximate knowledge of the 
interactions between hyperon(s) and nucleon(s), that can be
derived only from the few existing experimental data on
hyperon-nucleon scattering or from reactions involving hypernuclei.
New and precise experimental inputs are therefore awaited 
to provide reliable constraints on several 
parameters of the models, still loosely defined. 
This will open the possibility to improve the description of dense 
stellar objects able to reproduce the most recent astrophysical measurements 
of their mass.

In addition, the processes of $K^-$ absorption by two or few nucleons 
could be at the basis of the possible formation of kaon-nuclear bound states,
assumed to be
aggregates of nucleons strongly bound by an antikaon.
The existence of such states was predicted some years ago \cite{re:yama} and
over the last years several experiments have been pursuing their 
search. Some observations were reported for a bound $K^- pp$ system 
decaying in $\Lambda p$
\cite{re:finudaprl,re:ox,re:disto,re:fopi,re:FINUDAnew}, but they are still
awaiting  confirmation. Even if its observability could be questioned
as a consequence of 
the theoretically expected large widths \cite{re:oset}, 
its existence is not ruled out yet, as
well as those of different isospin partners 
that could be observed, for instance,
in $K^-$ absorptions producing $\Sigma N$ pairs.

This paper presents 
a study of the $K^-_{stop} A\rightarrow \Sigma^- p A'$ reaction
on $^6$Li, $^7$Li, $^9$Be, 
$^{13}$C and $^{16}$O, which enriches the existing meager 
set of available data. 
The data were collected in 2006-2007 by the FINUDA experiment,
a magnetic spectrometer
installed at the $e^+e^-$ DA$\Phi$NE collider, Laboratori Nazionali dell'INFN
di Frascati (LNF), Italy.   

This study follows a first analysis of kaon absorption by
one nucleon producing $\Sigma^\pm \pi^\mp$ final states \cite{re:sigmapi},
on the mentioned nuclear species. 
In the present analysis final states with one neutron, one negative pion and one proton only are
selected. The $\pi^-$ and the neutron can 
come from the $\Sigma^-$ hyperon decay 
($\Sigma^-\rightarrow n\pi^-$, B.R.=99.85\%), the proton can be
emitted promptly in the elementary  $K^-_{stop}(np)\rightarrow \Sigma^-p$ two-nucleon absorption
on a quasi-deuteron embedded in the $A$ nucleus. 
Hyperon ($Y$) and nucleon ($N$) pairs can also be emitted inclusively,
together with other particles, in many-nucleon absorptions.
The absorption mechanism is
expected to follow the ``Quasi-Free'' (QF) pattern \cite{re:heusi}: 
the kaon interacts at rest with a
nucleon pair (or cluster) having a Fermi momentum typical of the target nucleus \cite{re:fermi}
and a $YN$ pair is subsequently emitted; the  spectator nucleus recoils with
opposite momentum, in the nucleus center of mass. 
Since the kaons interact at rest
and close to the nuclear surface \cite{re:katz}, it is 
unlikely for the $K^-$ to be absorbed by the whole nucleus with a
following isotropic phase space emission of $\Sigma^-, p$ and $A'$ \cite{re:heusi}, so this
possibility has been discarded throughout this work. 

An earlier semi-inclusive analysis was performed by FINUDA on a first set of data,
in which the $\pi^-$ and the proton only were detected \cite{re:finuda1}. 
The $\Sigma^-p$ emission rate for $K^-$ induced interaction on
$^6$Li was assessed, with evidence that the signal 
observed in the proton momentum distribution
was a signature
of a two-nucleon QF absorption on $^6$Li.  A less clean signature was obtained in
a nucleus as heavy as $^{12}$C, due to a stronger dilution
effect caused by Final State Interactions
(FSI in the following).
The experimental signature of a two-nucleon QF kaon absorption is given by
a hyperon-nucleon pair emitted both with a momentum 
typically larger than 400~MeV/$c$. 
The particles from the $\Sigma^-$ decay have a continuous momentum distribution
that in FINUDA, for negative pions, extends 
from $\sim 80$ to $\sim 350$~MeV/$c$, and overlaps 
thus completely the momentum region of
$\pi^-$'s from hypernuclear formation.

This paper is organized as follows. In Sec. \ref{sec:exp} a short account
of the experimental set-up, already fully sketched out elsewhere 
\cite{re:apparato}, 
is given; in Sec. \ref{sec:dataselection} the data selection criteria are 
described. 
In Sec. \ref{sec:analysis} a study of the features
of some experimental spectra is presented. 
$^6$Li is chosen as reference nucleus,
due to its relatively simple structure and the cleanest signatures it
can provide. $^6$Li, 
as known from pion absorption experiments \cite{re:li6pioni},
can in fact be understood
as formed by a ``quasi''-$\alpha$ nucleus together with a $p+n$ pair or a
loosely bound ``quasi''-deuteron. The Fermi momentum of the $^6$Li 
subclusters can be modeled with fair accuracy \cite{re:yama6li}, so precise studies of
quasi-free absorptions are possible. In addition, with a nucleus as light
as $^6$Li the effects of FSI of the emitted particles with the
residual nucleus are limited.

The study of the distribution of the missing mass between the initial state and the
measured particles shows that some contributions of different QF reactions 
producing
$\Sigma^- p$ pairs, alone or together with other undetected particles, 
can be distinguished.
Emission rates of QF 
$\Sigma^-p$ reactions can be evaluated, per $K^{-}_{stop}$, for the
mentioned $p$-shell nuclei. The followed procedure is described 
in Sec. \ref{sec:ratesSection}. For the sake of conciseness, throughout the paper
we will indicate as ``semi-inclusive'' the emission of $\Sigma^-p$ pairs
recoiling against a nuclear system in its minimal mass configuration 
(coinciding with the ground state for stable nuclei, within
the experimental resolution), while the
notation ``inclusive''
will refer to the generic $A(K^-_{stop}, \Sigma^-pX)A'$ reaction, where $X$ can be 
any particle escaping detection, and $A^\prime$ the recoiling nucleus
in whichever energetic configuration. Capture rates are evaluated for both 
cases.

The obtained results are reported in Sec. \ref{sec:ratesSection}, 
while the
discussion of the results with a comparison with the few existing
data and the conclusions follow in Sec. \ref{sec:disc}.

\section{\label{sec:exp}Experimental setup}

The DA$\Phi$NE $e^+e^-$ collider at LNF
 provides $\phi(1020)$ mesons with a
luminosity of some $10^{32}$ cm$^{-2}$s$^{-1}$. The $\phi(1020)$,
produced almost at rest, decays 
about half
of the times into a pair of slow ($16.1\pm 1.5$~MeV), almost 
back-to-back charged kaons. A slight asymmetry in the $\phi(1020)$ decay is 
due to a small crossing angle between the colliding beams, which gives the
$\phi(1020)$   a small boost. 

FINUDA studied the reactions induced by the charged kaons.
In particular, the $K^-$
could be stopped in a set of eight thin targets ($\sim 0.25$ g/cm$^2$),
arranged coaxially as tiles around the beam and composed of materials
of pure isotopic composition 
with a mass number chosen
in the range $6 \le A \le 51$. In the years 2003--2007, 
in two runs, data corresponding to about
1.2 fb$^{-1}$ of $(e^+e^-)$ collisions at the $\phi$ energy 
were collected. The apparatus consisted of a
magnetic spectrometer with cylindrical symmetry, of 125 cm radius and 
255 cm maximum length, immersed in a 1 T solenoidal magnetic field
provided by a superconducting magnet and 
uniform to better than a few percent. 

The apparatus was 
able to detect the kaons before they reached the targets
by means of a thin scintillator hodoscope  
surrounding the beam pipe (named TOFINO in short) \cite{re:tofino}, and
an inner vertex detector composed of one layer of eight double-sided 
silicon microstrip modules (named ISIM) \cite{re:isim}. 
A dedicated algorithm, with 
a maximum inefficiency of $\sim 2\%$ (modulated 
depending on the target position), 
was developed to correctly identify the 
charge and momentum of the kaon before impinging on the target;
it exploited the available geometric and kinematic information.
Beyond the target array, the tracking 
of the charged particles emitted in the negative kaon interactions 
was performed; a momentum resolution as good as 0.6\% FWHM was achieved.
The tracking detectors stack consisted of 
one layer of ten double-sided
silicon microstrip modules (called OSIM) \cite{re:isim}, two layers of eight planar low-mass drift 
chambers (LMDCs) at a distance, respectively, of about 37 and 65 cm from the beam axis
\cite{re:lmd}, and a six-layer array of stereo-staggered  straw tubes \cite{re:straw}.
The tracking detectors were hold by an Aluminum clepsydra-shaped frame. 
The whole tracking region was filled and fluxed with Helium gas to minimize  
multiple scattering of the emitted particles.
The silicon detectors
and the drift chambers allowed also for the particle  
identification by means of the energy lost
in their active volume. Along with the time-of-flight (t.o.f.) information,
light minimum ionizing particles could be separated from protons 
with an identification efficiency
as large as 98\%.  
Deuterons and tritons could also be observed and 
identified, respectively, above 300 \cite{re:LambdaD} and 430~MeV/$c$ 
\cite{re:LambdaT}. 
The t.o.f. information was provided by a system composed of the
TOFINO scintillator array  as start detector,
and a stop scintillator barrel, 10 cm thick, located outside the 
tracking region and
facing the magnet coil (named TOFONE) \cite{re:tofone}.
The system delivered the trigger signal to the detector, and TOFONE also
allowed the detection of neutrons with an efficiency of about 10\%.

\section{\label{sec:dataselection}Data selection}

The analysis presented in this paper is performed on the data collected in the
second of the two FINUDA data takings (966 pb$^{-1}$). In this 
run  $p$-shell nuclei targets were used (solid: 2$\times ^6$Li, 2$\times ^7$Li, 2$\times ^9$Be; 
powder: $^{13}$C; liquid: $^{16}$O, in form of deuterated water contained
in an Aluminum-polycarbonate film envelop). The trigger thresholds
set in the first data taking did not allow neutrons with
energy below 11~MeV to be detected \cite{re:tesiBarbara}; therefore
these data cannot be effectively used in the present analysis, as too few 
useful $\Sigma^-$'s 
would be available after reconstruction.

$(n\pi^- p)$ events are selected requiring the proton and 
the incoming $K^-$ to
form the same vertex. The spatial resolution
of the $K^-$ stop point coordinates is 700 $\mu$m, evaluated 
through an extrapolation of the incoming track to the
target by means of the GEANE package \cite{re:geane}.
A cut on the distance of minimum
approach of the $\pi^-$ and the vertex in the target is applied 
tailored to the target
thickness and its spatial location, to allow the reconstruction  of
a $\Sigma^-$ secondary decay vertex. The minimum distance is 
required to be 1 to 1.2 cm.
High quality tracks ($\chi^2$ out 
of the track fitting procedure
corresponding to a probability of correctly
reconstructing a track larger than
$95\%$) are selected for both $\pi^-$ and proton.
Tracks hitting the Aluminum frame supports and not long enough to
reach at least the first drift chamber layer are discarded.

The momentum resolution of charged particles is not crucial for
this analysis. As reference, for 
$\pi^+$'s of 184.5~MeV/$c$ from the $\Sigma^+$ decay
at rest the FINUDA momentum resolution 
was 1\% \cite{re:sigmapi}; it worsened for lower momenta due to the smaller number of points
available for tracking, while it improved up to 
0.6\% for momenta of about 240~MeV/$c$, as 
measured from $K_{\mu 2}$ decays \cite{re:apparato}. 
From the same $\Sigma^+$ decay, 
the momentum resolution of neutrons
was assessed to be 5\% \cite{re:sigmapi}. 
We recall that the detection of neutrons is performed by means of a time of
flight measurement on a base of flight of 2 m at most. 

Neutrons in FINUDA are heavily
contaminated by the huge background
of $\gamma$'s from various particles' decays. In addition, some of them
are subject to rescattering before being detected by TOFONE. The
cut on the neutron speed applied in the analyses of Ref. \cite{re:sigmapi,re:nmneutrons} 
is partially released (1/$\beta >1$), 
to increase the neutron acceptance at all momenta. However, 
in this way a 
large number of $\gamma$'s and fake neutrons leaks
in the sample. Neutrons are identified when isolated TOFONE slabs 
(or pairs thereof) are found not connectable
to any charged track in the spectrometer, or far from spurious signals in the 
straw array that could derive from particles backscattered by the magnet yoke. 
An upper cut to 1/$\beta$ ($< 12$) is applied to reduce this
effect.
In case of multiple neutron candidates per event,  
their identification quality follows their energy deposit
on TOFONE slabs. In the present analysis, events are accepted if one 
neutron only is identified.

The background from fake neutral particles and from other contaminating 
QF reactions which can stem from the same final state
can be effectively reduced only after the application of proper kinematic cuts to the data.
In fact, the reconstruction efficiency for all neutral particles identified 
in FINUDA by t.o.f. is barely similar (averaged over the eight targets:
$(3.50\pm 0.01)\times 10^{-2}$ for neutrons, $(2.16\pm 0.01)\times 10^{-2}$ for 
$\gamma$'s,$(2.33\pm 0.01)\times 10^{-2}$ for $\pi^0$'s); therefore,
the possibility to discard contaminations from neutral particles other than 
neutrons mainly relies on a 
kinematic identification  of the searched reaction.

The main physical contribution to the $(n\pi^- p)$ sample is 
given by the $\Lambda n$ final state,
produced through a direct QF two-body absorption at the level 
of a few percent \cite{re:katz},
or via a $\Sigma$-$\Lambda$ conversion reaction, in both cases
with the $\Lambda$ decaying in $(\pi^- p)$ pairs (B.R. = 64\%).
In the present analysis $\Lambda$'s in the final state are eliminated by
means of a cut on the invariant mass of the detected $(\pi^-p)$ pair and 
on their angular correlation; in fact, the $\Lambda$ in this QF reaction
is produced with high momentum,
thus the particles from its decay are mostly forward emitted. To this purpose,
events are rejected if with $\cos\theta_{p\pi^-} > 0.6$.
The loss of $\Sigma^-$ events due to the application of these cuts is negligible.

\begin{figure*}[bHt]
\begin{center}
\includegraphics[width=\textwidth] 
{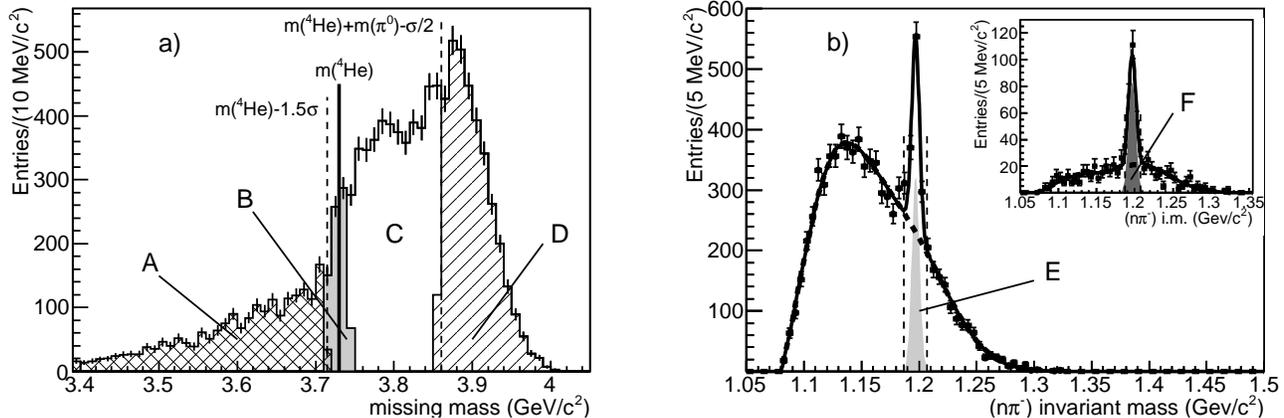}
\end{center}
\caption{a) Missing mass of the $K^-_{stop}\;^6\mathrm{Li}
\rightarrow n\pi^-p A'$ reaction:
the vertical lines mark the $^4$He mass value (solid), and the lower
limits set for the $K^-_{stop}\;^6\mathrm{Li}\rightarrow (\Sigma^-p) ^4\mathrm{He}$ and
$K^-_{stop}\;^6\mathrm{Li}\rightarrow (\Sigma^-p\pi^0)^4\mathrm{He}$ (dashed) 
reaction selection. The 
cross-hatched area (A) is filled by unphysical events, the grey one (B) by
events
from the QF $K^-_{stop}\;^6\mathrm{Li}\rightarrow (n\pi^-p) ^4\mathrm{He}$ reaction,
the hatched one (D) by 
events compatible with the presence of at least one undetected
$\pi^0$. b) Invariant mass of the $(n\pi^-)$ system for the
$K^-_{stop}\;^6\mathrm{Li}\rightarrow n\pi^-p A'$ reaction: in the inset events selected in
the grey area (B) of a)  are selected. The grey-filled gaussian areas correspond to the
$\Sigma^-$ signals, see text for details. The regions
denoted as E and F and delimited by dashed vertical lines indicate, in the two plots, the chosen mass window for the 
evaluation the $\Sigma^-$ signal integral.
}
\label{fig:missingTotal}      
\end{figure*}

A second large contamination comes from the QF one-nucleon  
$K^-p \rightarrow \Sigma^+\pi^-$ reaction, followed by either  
$\Sigma^+\rightarrow p\pi^0$ with one of the $\gamma$'s from $\pi^0$
mimicking a neutron (or by $\Sigma^+\rightarrow n\pi^+$ with a
$\pi^+/p$ misidentification, whose occurrence is however thoroughly suppressed 
thanks to the apparatus PID capabilities), or by a
$\Sigma^+ n\rightarrow \Lambda p$ conversion 
and the following $\Lambda\rightarrow n\pi^0$ decay (B.R. = 36\%). In the latter case, 
a $\pi^0$ is missing. The $\Sigma^+\pi^-$ QF absorption on one nucleon occurs
with an emission rate of about 18\%/$K^-_{stop}$ in $^6$Li 
\cite{re:sigmapi}, so the contamination of this channel can be quite sizeable.
The largest part of these events can be rejected applying a cut on the 
secondary vertex distance and, eventually, on the missing mass of the 
reaction. This quantity, defined as
\begin{equation}
\mathcal{M} = \sqrt{(M_A-E_n-E_{\pi^-}-E_p)^2 -(\vec p_n+\vec p_{\pi^-} +\vec p_p)^2}
\end{equation}
where $M_A$ is the mass of the target nucleus, and E and $\vec p$ the 
energy and momentum
vector of the three measured particles, is shown in
Fig. \ref{fig:missingTotal}a) for the 
$^6\mathrm{Li}(K^-_{stop}, n\pi^-p)\mathrm{A}'$ reaction, after the cuts for the
mentioned contaminations.
$^6\mathrm{Li}$ is shown as typical case; spectra from other targets are
similar. For ease of observation,
the spectra in Fig. \ref{fig:missingTotal} are not corrected by 
the apparatus acceptance, that will be accounted for in the following.

A large part of events in the missing mass distribution,  
to the left
of the mass value of the ground state of the 
recoiling nucleus (in the case of $^6$Li:
$^4$He mass, 3.73~GeV/$c^2$), is affected by a wrong neutron momentum
evaluation or misidentification. They mainly
derive from the mentioned $\gamma$-neutron
misidentification and from neutrons scattered 
before detection, to which a larger momentum is incorrectly assigned.
A large source of photons is the mentioned
single nucleon absorption in $\Sigma^+\pi^-$, with $\Sigma^+\rightarrow p\pi^0$ 
(B.R. = 51.57\%); as well, all QF reactions involving
$\Sigma^0$'s produce $\gamma$'s.
Also a small number of physical $\Sigma^- p$ events 
(less than 10\% of the total semi-inclusive sample) belongs to this region, 
those in which more than one nucleon is emitted in the absorption. 
Since events with one neutron only are accepted, a misidentification 
between the prompt and the $\Sigma^-$ decay neutron may have occurred
in this case.

For the sake of brevity the recoiling configuration
with minimum mass will be denoted
in the following as $A'_{g.s.}$, as a shortcut valid also for the cases in which the
recoiling system is unstable (like, {\it e.g.}, $^5$He). 
Due to the missing mass resolution, which
varies in the range $(9-13)$ MeV/$c^2$ according to the target location and
composition, the notation $A'_{g.s.}$ thus includes also fragmented recoiling 
nuclear systems,
produced in the absorption on three nucleons or more, which cannot be 
distinguished from the ground state.

In Fig. \ref{fig:missingTotal}a)
the vertical solid line marks the $A' \equiv ^4$He mass, and the dashed 
one on the left 
the lower limit chosen for the analysis.
The events in the cross-hatched area, indicated in the plot as A,
are removed from the studied sample. Monte Carlo evaluations show that 
this cut rejects just a negligible number of physical $\Sigma^- p$ events.

The $(n\pi^-)$ invariant mass of the remaining events for the $^6$Li target 
is displayed in Fig. \ref{fig:missingTotal}b). 
The central value of the $\Sigma^-$ signal is at
$(1197.0\pm 3.4 (\sigma))$~MeV/$c^2$, from a fit
by means of a gaussian function
superimposed to a fifth order Chebychev polynomial. The integral of the
peak within $\pm 2\sigma$ and of the polynomial background ($B$)
in the same
mass interval (the region marked as E in Fig. \ref{fig:missingTotal}b) gives 
the signal ($S$) sensitivity $\mathscr{S} = S/\sqrt{S+B} = 15.48$,
corresponding to a signal/background ratio $S/B = 0.84$. 

In the missing mass distribution displayed in Fig. \ref{fig:missingTotal}a) 
at least two parts may be singled out which can be addressed as signatures of
distinct QF two-nucleon $K^-$ absorptions. Around the $^4$He mass value, a small
enhancement, two bins wide, can be observed: it is due to the   
$K^- \;^6\mathrm{Li}\rightarrow\Sigma^- p ^4\mathrm{He}$ two-body QF
reaction (grey solid area in the plot, labeled as B), plus 
many-nucleon absorptions in which the four-nucleon 
recoiling system mass falls within the missing mass selected range.
The extension of the grey area,
corresponding to three times the missing mass experimental resolution,
is chosen to minimize the contributions of these many-nucleon 
absorptions.
Depending on the
target nucleus, the suppression of the overlapping reactions
is, though, 
only partly successful. A $3\sigma$
range allows to discard the completely fragmented recoiling configurations, for all the studied nuclei.
Conversely, configurations in which one or two more nucleons are emitted together with the remainder of the
nuclear system can be separated from the minimal mass configuration with an
efficiency that is
usually larger for heavier nuclei. 
Since the relative occurrence of each of these reactions
is unknown, in the following an integrated rate only will be quoted.

A second enhancement can be seen at higher missing mass where
the threshold for  $\pi$ production opens:
the hatched grey area in the plot (D) corresponds to events 
compatible with the
$K^- \; ^6\mathrm{Li}\rightarrow \Sigma^- p \pi^0\; A'$
QF reaction, in which the
$\pi^0$ escapes detection. 

These two classes of reactions alone
cannot explain the central bulk of the spectrum, indicated in the
figure as C (open area). This area may be filled by contaminating
background reactions which feed the $(np\pi^-)$ final state as well as
by QF $\Sigma^-p$ two nucleon
absorptions (recoiling against nuclear systems in different energy
configurations) possibly followed by FSI of the emitted particles, 
or by many-nucleon absorptions 
with the emission of $\Sigma^-,\, p$ and additional, undetected, nucleons and pions.

A quantitative study of the most likely contaminating reactions 
leaking through the data selection
criteria has been performed by means of Monte Carlo simulations.
Table \ref{tab:contam} reports the values of these contaminations per
generated event, 
averaged over the eight available targets: 
the simulated events are filtered through the reconstruction and
analysis chain 
and selected in the 
$\Sigma^-$ invariant mass band (E region). 

\begin{table*}[Hbt]
\caption{Average contaminations per event
(over eight targets) of physical background 
reactions, for simulated events subject to the described analysis cuts and
selected in the $\Sigma^-$ invariant mass region. For the sake of comparison, the last line
reports the reconstruction
efficiency for the signal reaction $K^-_{stop}(np)\; [A-np]\rightarrow \Sigma^- p\;  [A-np]$.}
\label{tab:contam}
\begin{ruledtabular}
\begin{tabular}{lc}
reaction & contamination ($\times 10^{-7}$/event) \\ \hline
$K^-_{stop}p\ [A-p]\rightarrow \Sigma^+ \pi^-\; [A-p]$ & $120.5 \pm 3.3$ \\
$K^-_{stop}np\; [A-np]\rightarrow \Sigma^0 n\; [A-np],\; \Sigma^0n\rightarrow\Lambda n$ & $30.1 \pm 1.3$ \\
$K^-_{stop}p\ [A-p]\rightarrow \Sigma^- \pi^+\; [A-p],\; \pi^+ n\rightarrow \pi^0 p$ & $11.8 \pm 1.0$ \\
$K^-_{stop}np\; [A-np]\rightarrow \Sigma^+ \pi^- n\; [A-np]$ & $8.8 \pm 0.7$ \\
$K^-_{stop}np\; [A-np]\rightarrow \Sigma^0 n\; [A-np]$ & $6.7 \pm 1.0$ \\
$K^-_{stop}np\; [A-np]\rightarrow \Lambda\pi^0 n\; [A-np]$ & $4.8 \pm 0.6$\\
$K^-_{stop}p\; [A-p]\rightarrow \Sigma^0 \pi^0\; [A-p]$ & $4.6 \pm 0.5$ \\
$K^-_{stop}np\; [A-np]\rightarrow \Lambda n\; [A-np]$ & $2.4 \pm 0.3$ \\
$K^-_{stop}nn\; [A-nn]\rightarrow \Sigma^- n\; [A-nn],\;  \Sigma^- p\rightarrow\Lambda n$ & $2.3 \pm 0.4$\\
\hline
$K^-_{stop}np\; [A-np]\rightarrow \Sigma^- p\;  [A-np]$ & $ 9469.8\pm 30.0$ \\
\end{tabular}
\end{ruledtabular}
\end{table*}

According to the position of the
target, different sectors of the spectrometer are spanned by the tracks, with different detector
efficiencies; the error quoted in Tab. \ref{tab:contam} is systematic and
takes into account their maximum spread. The statistical error, in comparison,
is negligible.

The main contaminating contribution, on the order of 
$10^{-5}$/event, is given 
by the one-nucleon $K^-p\rightarrow\Sigma^+\pi^-$
absorption; the contamination of the other reactions to
the selected sample is at least one order of magnitude less. 
For the sake of comparison, the reconstruction efficiency for the 
$K^-_{stop}(np)\; [A-np]\rightarrow \Sigma^- p\; [A-np]$ QF
reaction is reported in the last line of the table.

\begin{figure}[Hbt]
\includegraphics[width=0.5\textwidth]{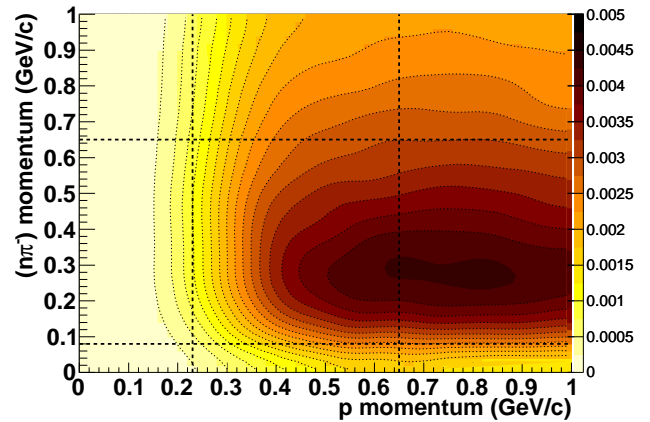} \\
\caption{(color online) 2-dimensional projection of the acceptance 
multidimensional map for
$n\pi^-p$ events emitted in $K^-$ induced reactions, integrated over the
full FINUDA target volumes.}
\label{fig:acceptance}      
\end{figure}

The experimental spectra displayed in the following are corrected for the apparatus acceptance.
The acceptance correction is made per-event and is determined 
through the generation of about $3\times 10^9$ 
$K^-_{stop}A\rightarrow n\pi^- p A'$ events with uniform distribution of the final
particle momenta over a wide kinematic range, exceeding that allowed 
by the QF reactions under study.
The acceptance function is determined in each cell of the multidimensional array containing the
kinematic coordinates of each of the three particles (momenta and production
vertices); it is defined as the ratio between the number 
of events surviving the reconstruction, and the total 
number of generated events. 
It embeds the reconstruction efficiency as well as the geometric
acceptance of the apparatus, while distortions due to 
inefficiencies of tracking detectors
are not included. In the experimental plots shown in the
following the bin errors account for 
the systematic uncertainty of the acceptance correction and the statistical 
uncertainty of the experimental bin, added in quadrature.
A 2-dimensional projection of the multidimensional acceptance map is shown in Fig. 
\ref{fig:acceptance}, integrated over the full targets fiducial volume, as a 
contour plot of the momentum of the 
$(\pi^-n)$ pair versus the proton momentum.

The 2-dimensional projection of the acceptance shows a moderate increase
in the ~200-500~MeV/$c$ $(n\pi^-)$ pair momentum range.
In the figure the dashed lines indicate the kinematic limits chosen for the
analysis. Outside these limits the acceptance is 
small and its uncertainty large, consequently the correction  would
affect the experimental spectra with large systematic errors.
The chosen ranges for the application of the acceptance correction are $80-650$~MeV/$c$ for 
the $(\pi^- n)$ pair, and $230-650$~MeV/$c$ for $p$. 
These ranges cover completely the $\Sigma^-p$ QF production kinematics.

\section{\label{sec:analysis} Study of the QF $K^- A\rightarrow \Sigma^- p\; [A-(np)]_{g.s.}$ reaction}

The invariant mass of the $(n\pi^-)$ pair for events selected in the B 
area of Fig. \ref{fig:missingTotal}a) is reported in the inset of Fig.
\ref{fig:missingTotal}b).
A similar fit with a gaussian function and a fifth
order Chebychev polynomial reports a large improvement of the $S/B$ ratio,
that becomes 2.38 for the $^6\mathrm{Li}$ sample. 
Background subtracted data in the signal region F (evidenced in the picture
in darker grey shade) 
are counted to evaluate the emission rates of the semi-inclusive QF 
$\Sigma^-p$ reaction. 

\begin{figure*}[Hbt]
\begin{center}
\includegraphics[width=\textwidth]{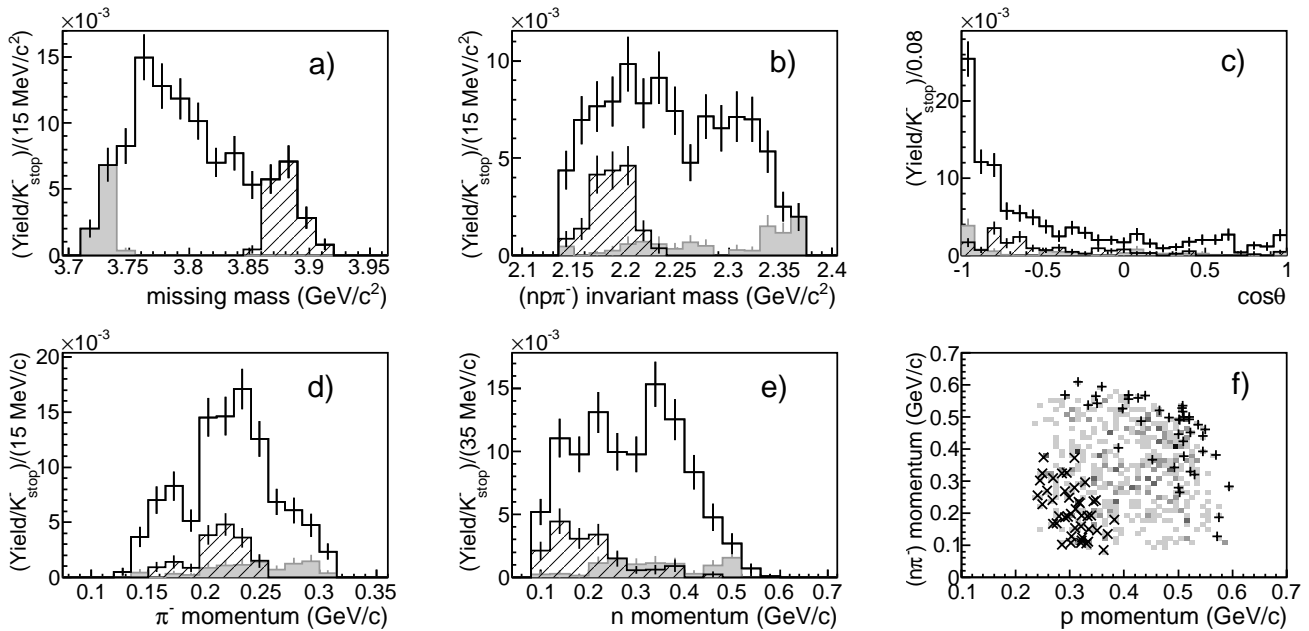}
\end{center}
\caption{Acceptance corrected distributions of some observables
measured in the $K^-_{stop}\;^6\mathrm{Li}\rightarrow n\pi^-p A'$
reaction, with the invariant mass of the $(n\pi^-)$ pair selected
in a 12~MeV/$c^2$ mass window centered on the $\Sigma^-$ mass value.
a) Missing mass of the reaction, for events selected in the E
region of the total missing mass plot of Fig. \ref{fig:missingTotal}b). 
The grey distribution corresponds to events belonging to region F only,
the hatched one to events from region D of Fig. \ref{fig:missingTotal}a)
after the selection in the $\Sigma^-$ mass window (G).  
b) Invariant mass distribution
of the $(n\pi^-)p$ system, same selections. c) Distribution of the
angle between the $(n\pi^-)$ pair and the proton, same selections.
d)-e): Momentum distribution of the $\pi^-$ and the neutron, same
selections. f): Scatter plot of the $(n\pi^-)$ system vs proton momentum.
Events belonging to region F (upper right part, plus markers) and 
to the G sample (lower left part, cross markers), 
are superimposed.
}
\label{fig:6LiPlots}      
\end{figure*}

Fig. \ref{fig:6LiPlots} reports, for the $^6$Li targets, 
several acceptance corrected spectra obtained selecting events in a 
$\pm 6$~MeV/$c^2$ wide window 
centered on the mean value of the $\Sigma^-$ peak (E area 
in Fig. \ref{fig:missingTotal}b)). 
The plots report, respectively, (a) the missing mass distribution, (b) the invariant
mass of the  $(n\pi^-)p$ system, 
(c) the distribution of the angle between the
$(n\pi^-)$ system and the proton, the momenta distributions of (d) the $\pi^-$ 
and (e) the neutron, and the scatter plot of the momentum
of the $(n\pi^-)$ pair versus the proton momentum. 
The area of the plots is normalized to the
total production rate of inclusive $\Sigma^-p$ final state described
in the following section, which includes the background contribution.

In each a)-d) plot, the grey parts correspond
to events selected in the F region of the $(n\pi^-)$ invariant mass plot 
shown in Fig. \ref{fig:missingTotal}b), while the 
hatched region corresponds to the subset of events located in the D region 
and selected in the mentioned $\Sigma^-$ mass window
(this data set will be called as G in the following as a shortcut). 

In the missing mass spectrum of Fig. \ref{fig:6LiPlots}a) 
the contributions from F and G samples are clearly seen. In the 
$(n\pi^-)p$ invariant mass distribution (b), the events from the 
semi-inclusive QF $\Sigma^- p$ reaction (F) tend to cluster in the
region around 2.35~GeV/$c^2$. This correspond to the threshold of
the available phase space, while events compatible with an additional pion
are located in a lower mass region. We recall that some
of these events could also come from the $\Sigma^-\pi^0$
decay of $\Sigma^-(1385)$.

Fig. \ref{fig:6LiPlots}c) shows
the distribution of the angle between the $(n\pi^-)$ pair and the proton.
This distribution indicates 
that the events in the $\Sigma^-$ band have a back-to-back correlation
 (grey histogram), 
as expected in a two-body Quasi-Free absorption. On the other hand,
the events selected in the hatched area (G)
do not exhibit any favored topology, as typical of a three-body reaction.
 
Fig. \ref{fig:6LiPlots}d) and e) report, respectively, the momentum 
distributions for the $\pi^-$ and the neutron. No peculiar enhancement is
observed in the distributions. The events selected in the $\Sigma^-$
mass window distribute continuosly in a similar way for both the decay
particles
(with a larger background contamination for neutrons),
indicating that both are most likely coming from the decay in flight of the 
$\Sigma^-$ \cite{re:finuda1}. 

Finally, in Fig. \ref{fig:6LiPlots}f) the events from regions F and
G are superimposed to the full scatter plot 
with different markers (pluses for F set, crosses for G set): 
the distribution in the upper right part of the pad
corresponds to the events of set F, the one in the lower left corner 
to events of set G. Set F events prefer large momenta for both the
$(n\pi^-)$ system and the proton, a typical feature of $YN$ pairs produced
in two-nucleon absorption, as mentioned earlier.
 
\begin{figure*}[Hbt]
\begin{center}
\includegraphics[width=\textwidth]{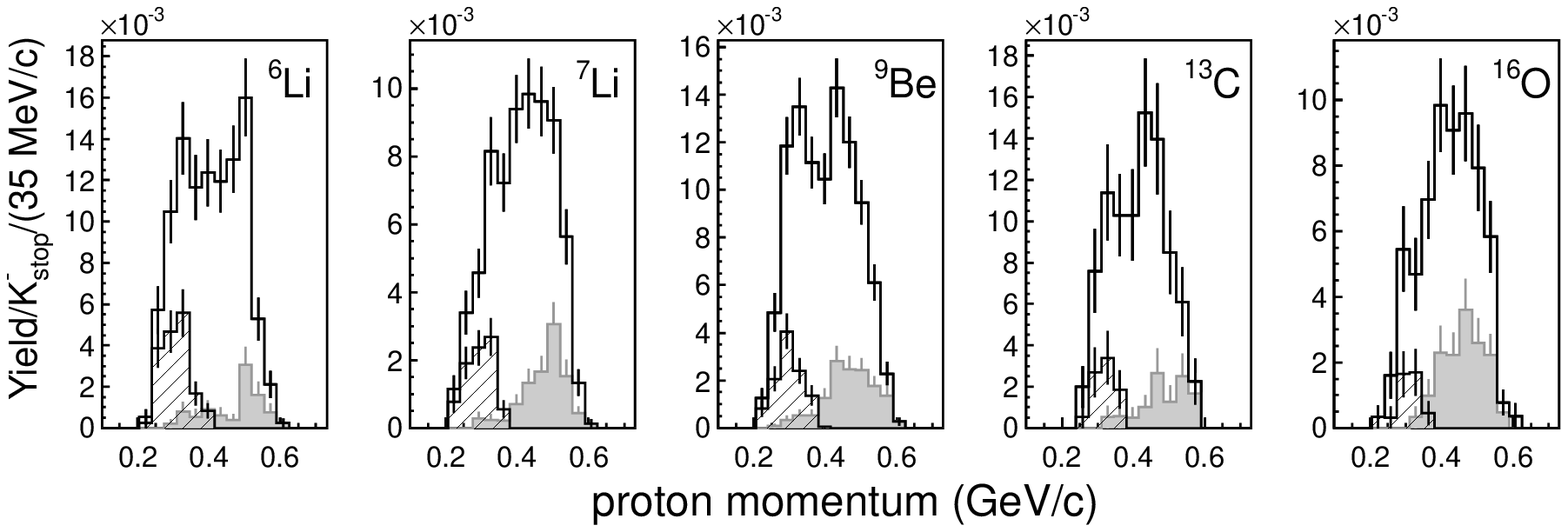} \\
\includegraphics[width=\textwidth]{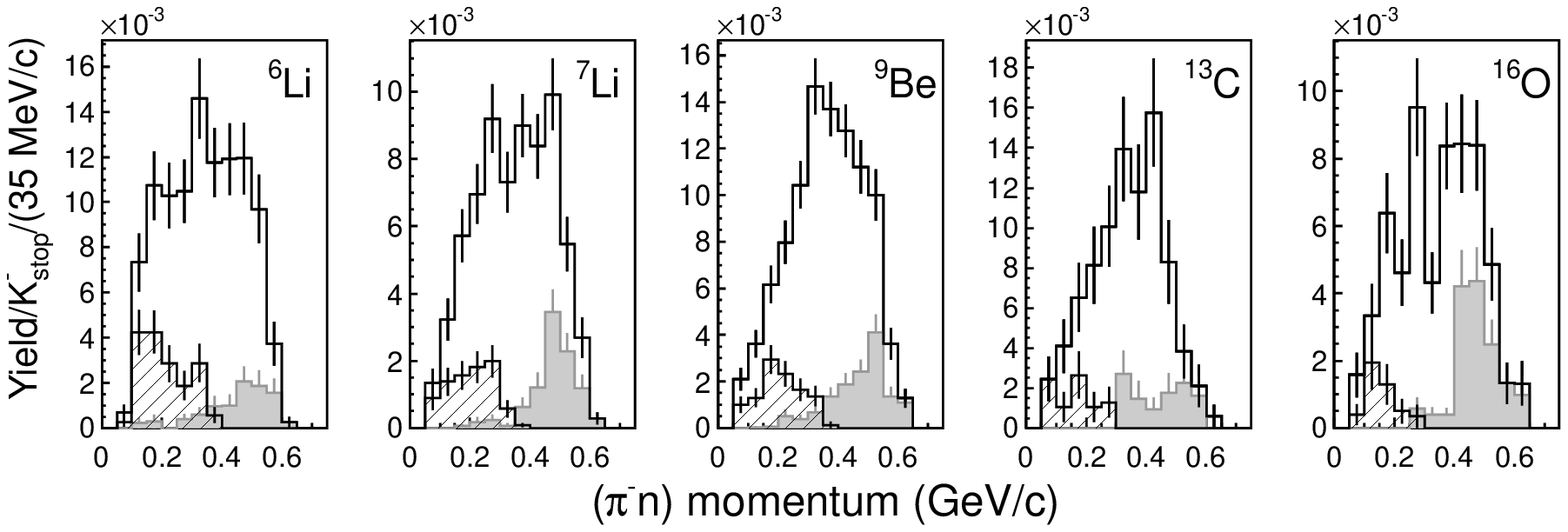}
\end{center}
\caption{Upper row: acceptance corrected proton momentum spectrum
for all targets, for events selected in the E region of Fig.
\ref{fig:missingTotal}b). The events in the grey area come from region
F only, those in the hatched area from region G (see text for
description). Lower row: 
selections as described above, $(n\pi^-)$ momentum distribution 
in the $\Sigma^-$ mass region for all the targets. }
\label{fig:momenta-pSigma_allTargets}      
\end{figure*}

The acceptance corrected momentum distribution for prompt protons and for the
$(n\pi^-)$ system selected in the $\Sigma^-$ mass window 
for all the studied $p$-shell nuclei are
shown in Fig. \ref{fig:momenta-pSigma_allTargets}, normalized to the 
evaluated emission rates. 
In all of them, as in the previous pictures, the
grey area corresponds to events selected in a missing mass window 
centered on the minimum 
recoiling mass nucleus (F), while the hatched  
areas to events located above the $\Sigma^- p\pi^0$ threshold (G). 
The proton momentum distributions indicate that most of 
the events in F area belong to semi-inclusive QF $\Sigma^- p$ reactions, and are
characterized by momenta above 
500~MeV/$c$.
The same feature is shown by $\Sigma^-$'s, as
already displayed in the scatter plot of Fig. \ref{fig:6LiPlots}f) for $^6$Li.

A distinctive trait of the largest effect of FSI for
heaviest nuclei can be seen in the broadening of the QF prompt proton
signal as the target nucleus mass number increases, as already observed in
Ref. \cite{re:finuda1}.

\section{\label{sec:ratesSection} Emission rate evaluations}

For reactions including $\Sigma^-$'s, emission rates only can be assessed,
following the method already applied in several previous
FINUDA papers \cite{re:germanoplb,re:sigmapi,re:finuda1};
in this analysis no correction is applied for possible pion
attenuation effects nor for the $\Sigma^-$ loss due to $\Sigma$-$\Lambda$ 
undetected conversions.
Following Ref. \cite{re:sigmapi}, the evaluated rates are finally scaled by
the loss rate of slow $\Sigma^-$'s due to their 
single-nucleon capture in nuclei.
Two data samples have been examined: events belonging  
to the full $(n\pi^-)$ invariant mass spectrum
of Fig. \ref{fig:missingTotal}b) (region E), and events selected 
in the missing mass 
region corresponding
to the semi-inclusive QF $\Sigma^- p$ reaction (inset, region F -- we recall that by
``semi-inclusive'' the $\Sigma^- p$ production recoiling against a nuclear system
in its minimum energy configuration is understood).
The emission rate is defined, for each target, as
\begin{equation}
R_A = N_{\Sigma^- p}/N_{K^-_{stop}}.
\label{eq:ratio}
\end{equation}
The numerator gives the absolute number of $\Sigma^- p$ events, which corresponds to
the $N^C_{\Sigma^- p}$ events counted in the $\Sigma^-$ peak of the $(n\pi^-)$ 
spectrum after proper background
subtraction and corrected by the global
efficiency of the channel: $N_{\Sigma^- p} = N^C_{\Sigma^- p}/\epsilon_{\Sigma^-p}$. 
The global efficiency can be factorized in two parts. 
The first depends on the trigger efficiency and reconstruction performance 
(determined, for each reaction, by the FINUDA 
Monte Carlo). 
The second one depends on the detectors' efficiencies, 
which are local and whose effects are different according to 
the event topology (track curvatures, charges, lengths, 
number of hits in each track and number of tracks
in a given apparatus sector). 
As far as the t.o.f. and vertex detectors are concerned, their response
was completely digitized in the FINUDA simulation code, where all thresholds and
maps of inefficient channels had been listed: therefore, their instrumental inefficiency is taken into
account at the hit definition and the following track reconstruction levels. Consequently, the efficiency for neutron
detection is already embedded in the reconstruction.
To get information on the detector efficiencies in the
tracking volume
$K_{\mu 2}$ decay real events have been used for positive tracks, while for negative tracks 
the information
from Bhabha prongs as well as from the charged pions from $K^0_S$ decays 
has been exploited. 
As figures of merit, a typical integrated detector efficiency for a positive track of about 250~MeV/$c$
is (73--78)\%, while for negative tracks of similar momentum it spans the range (63--86)\%. The inefficiencies are mainly due to the dead zones 
corresponding to Aluminum supports, that were arranged
to maximize the acceptance for $\sim$270 MeV/$c$ negative pions from 
hypernuclear decays. 
The difference among these efficiencies in different apparatus regions 
is also due to the 
performances of OSIM and LMDC's.

This part of the global efficiency carries the largest sources 
of systematic uncertainty in the evaluation of the
emission rate, as will be reported in the following. 

In the denominator of eq. (\ref{eq:ratio}), the number of $K^-$'s stopped in each target is
given by the reconstructed number of vertices, after the elimination of those with a
fake $K^\pm$ assignment, scaled by the global $K^-_{stop}$ detection efficiency: 
$N_{K^-_{stop}} = N^{Rec}_{K^-_{stop}}/\epsilon_{K^-}$. This efficiency takes into account the
trigger acceptance (depending on the reaction), and also the instrumental efficiency of the
detectors delivering the trigger signal (TOFINO and TOFONE). Moreover, 
it includes 
the efficiency of the kaon stopping point determination procedure.
$\epsilon_{K^-}$ can be evaluated, target by target, by a full simulation 
in which fourteen types of 
QF reactions were injected (including rescattering and $\Sigma$-$\Lambda$
conversion effects), each one with rates 
extrapolated from the few available measurements. The simulated statistics 
correspond to 
about 1.3 million $K^-_{stop}$ events per target. The global 
$K^-_{stop}$ detection
efficiency varies in the range (22--46)\%, 
depending on the target; disfavored cases correspond to thicker targets  (in
particular, $^{13}$C) for which the GEANE extrapolation is more critical.

This rate evaluation procedure, whenever possible, is tested against
a second method based on the
coincidence count of the searched topologies 
and the $\mu^+$ from $K_{\mu 2}$ decay 
emitted from the opposite $K^+$ vertex, in the same $\phi$ decay 
(tagged events). 
Reactions with such a coincidence
occur with a full trigger acceptance, but they suffer for reduced 
statistics and therefore they bear a larger statistical uncertainty. The two methods, when
applied on samples of enough statistics (namely, on the inclusive 
sample E), 
lead to results in agreement within the statistical uncertainty. 
Their relative spread is taken into
account as a source of systematic error.

Table \ref{tab:emission} summarizes the results obtained for the rates evaluated for the E and
F experimental subsets. They refer to the full kinematic range of the reactions.
For each of the
studied nuclei, the $S/B$ value is reported (averaged, 
when two different targets of the same nuclear species
are available) along with the sensitivity $\mathscr{S}$, which
gives an estimation of the statistical significance of the observed signal.
The subsequent columns report 
the number of counted $\Sigma^- p$
events, and the value of the emission rate, with both statistical and 
systematic errors. The total systematic error is obtained by summing
in quadrature the contributions from different uncertainty sources,
described in the following.
When two targets are available, the weighted
average of the rates obtained for each of them is reported. 
The spread between the values 
obtained for targets of the same species represents the second
largest source of systematic 
error. This is especially true for the 
$^6$Li targets whose geometrical position (in the $\phi$(1020) 
anti-boost direction)
was unfavourable to stop kaons, being a larger part 
of them
stopped by ISIM before they could reach the targets.

\begin{table*}[Ht]
\caption{
$\Sigma^- p$ emission rates in $K^-_{stop}A$ absorption in several $p$-shell nuclei, listed in the
first column. The second column, from Ref. \cite{re:sigmapi}, reports the correction factor
for undetected $\Sigma^-$'s undergoing $\Sigma$-$\Lambda$ conversion at rest. 
The third, fourth, fifth and sixth 
column report, respectively, the $S/B$ ratio, the sensitivity $\mathscr{S}$, the number of observed
$\Sigma^- p$ events after background subtraction 
and the measured emission rate $R_A$, in units of $10^{-2}/K^-_{stop}$.
For each nuclear species, the first row refers to 
events belonging to the region marked as E in Fig. \ref{fig:missingTotal}b), the second one
to events selected in the F region of Fig. \ref{fig:missingTotal}b).
For all rates, statistical and systematic uncertainties are quoted. The 
systematic error has been evaluated adding in quadrature the contributions from
the systematic uncertainty sources described in the text.}
\label{tab:emission}
\begin{ruledtabular}
\begin{tabular}{lccccc}
$A$ & $\Sigma^-_{loss}\; (10^{-2})$ & S/B & $\mathscr{S}$ & $N^C_{\Sigma^- p}$ & $R_A\; (10^{-2}/K^-_{stop})$ \\ \hline
\multirow{2}{*}{$^6$Li} & \multirow{2}{*}{$30\pm 2$} & 0.84 & 15.48 & $828\pm 36$ &  $4.72 \pm 0.61_{stat} \pm 1.05_{syst}$ \\ 
                        &           & 2.38 & 11.57 & $224\pm 18$ & $1.20 \pm 0.13_{stat} \pm 0.27_{syst}$\\
\hline
\multirow{2}{*}{$^7$Li} & \multirow{2}{*}{$26\pm 2$} & 0.92 & 18.19 & $734\pm 42$ &  $3.33 \pm 0.42_{stat} \pm 0.52_{syst}$ \\
       &                            & 2.56 & 11.95 & $211\pm 18$ & $0.67 \pm 0.08_{stat} \pm 0.10_{syst}$\\
\hline
\multirow{2}{*}{$^9$Be} & \multirow{2}{*}{$38\pm 2$} & 0.92 & 17.22 & $983\pm 55$ &  $4.67 \pm 0.59_{stat} \pm 0.41_{syst}$ \\
              &                            & 3.07 & 11.97 & $227\pm 17$ & $0.88 \pm 0.09_{stat} \pm 0.03_{syst}$\\
\hline
\multirow{2}{*}{$^{13}$C} & \multirow{2}{*}{$12\pm 1$} & 0.90 & 11.43 & $395\pm 54$ & $4.15 \pm 0.78_{stat} \pm 0.52_{syst}$ \\
        &                            & 1.87  & 5.15  & $48\pm 9$ & $0.46 \pm 0.09_{stat} \pm 0.02_{syst}$ \\
\hline
\multirow{2}{*}{$^{16}$O} & \multirow{2}{*}{$36\pm 3$} & 1.21 & 12.36 & $348\pm 26$ & $3.42 \pm 0.65_{stat} \pm 0.45_{syst}$ \\
        &                            & 4.56 & 8.22 & $86\pm 17$ &  $0.71 \pm 0.13_{stat} \pm 0.05_{syst}$ \\
\end{tabular}
\end{ruledtabular}
\end{table*}

Table \ref{tab:syst} reports the relative systematic errors 
for the two data samples, 
coming from different sources. Since the errors vary depending on the target
composition and location, their maximum spread 
is quoted. In addition to the
mentioned errors due to $K^+/K^-$ identification
inversion, different methods for the capture rate
evaluation,
target pairs located in different acceptance regions and tracking and detector efficiencies, other sources are due
to changes in the selection criteria,
to the peak/background fitting procedure, fitting functions and algorithm, 
and to the $\Sigma^-$ window width chosen for the signal selection. 

Overall, a maximum systematic error of 22\% affects both the semi-inclusive 
and inclusive rates.

\begin{table*}[Hbt]
\caption{
Relative systematic errors of $\Sigma^-p$ emission rates 
from different 
sources, for the semi-inclusive $K^-_{stop}A\rightarrow \Sigma^-p A'_{g.s.}$ 
reaction ($R(A'_{g.s})$) and for the inclusive 
$K^-_{stop}A\rightarrow \Sigma^-p X A'$ one ($R(A')$): the two values indicate 
the range of variability, depending on the target composition and location.
}
\label{tab:syst}
\begin{ruledtabular}
\begin{tabular}{lcc}
{Syst. error source} & {$\Delta R(A'_{g.s.})/R(A'_{g.s.})$ (\%)} & 
{$\Delta R(A')/R(A')$ (\%)}\\ 
\hline
{$K^+/K^-$ id. inversion and ISIM efficiency} & 0.4--1.9 & 0.4--1.9 \\
{$\mu$ tag rate evaluation} & -- & 3--11  \\ 
{tracking and tracking detector efficiencies} & 0.9--7.8 & 0.4--6.2 \\
{targets in different acceptance regions} & 0--20 & 0--17 \\
{selection and signal extraction efficiency} & 0.7--3.8 & 0.3--10.0 \\ \hline
total & 4--22 & 9--22 \\
\end{tabular}
\end{ruledtabular}
\end{table*}

\section{\label{sec:disc} Discussion and conclusions}

\begin{figure}[Hbt]
\includegraphics[width=0.5\textwidth]{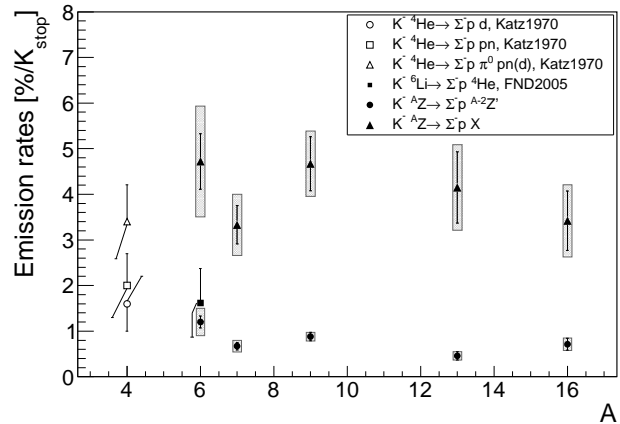}
\caption{
Emission rates of the 
$K^-\, A\rightarrow \Sigma^- p [A-(np)]$ semi-inclusive and
$K^-\, A\rightarrow \Sigma^- p X A^\prime$ inclusive
reactions
for several nuclei. Measurements on $^4$He are from Katz1970:\cite{re:katz},
measurements for $A\ge 6$ are by FINUDA (FND2005:\cite{re:finuda1} and present
work). For the measurements described
in the present paper, the 
error bars report the 
statistical uncertainty, while the 
grey boxes show the statistical and systematic error
added in quadrature; for the other measurements, the total error is reported.}
\label{fig:SPrates}      
\end{figure}

Fig. \ref{fig:SPrates} summarizes pictorially the 
available measurements of emission rates for the 
$K^-_{stop} A\rightarrow \Sigma^- p A'$ reaction: the present measurements 
(Tab. \ref{tab:emission}, full triangles and circles) 
add up to the few existing points on $^4$He \cite{re:katz} 
(open triangle, open 
square and open circle) and on
$^6$Li by the former FINUDA evaluations \cite{re:finuda1} (full square).
For the measurements in $^4$He, the open circle corresponds to
$R(K^-\ ^{4}\mathrm{He}\rightarrow \Sigma^-pd) = (1.6\pm 0.6)\%/K^{-}_{stop}$, 
the open square to  
$R(K^-\ ^{4}\mathrm{He}\rightarrow \Sigma^-ppn) = (2.0\pm 0.7)\%/K^{-}_{stop}$ 
and the open triangle to the sum of 
$R(K^-\ ^{4}\mathrm{He}\rightarrow \Sigma^-p\pi^0np) = 
(1.0\pm 0.4)\%/K^{-}_{stop}$, 
$R(K^-\ ^{4}\mathrm{He}\rightarrow \Sigma^-p\pi^0d) = 
(1.0\pm 0.5)\%/K^{-}_{stop}$ and 
$R(K^-\ ^{4}\mathrm{He}\rightarrow \Sigma^-p\pi^+nn) = 
(1.4\pm 0.5)\%/K^{-}_{stop}$.
The former semi-inclusive FINUDA measurement on $^6$Li was
$R(K^-\ ^{6}\mathrm{Li}\rightarrow \Sigma^-p [^6\mathrm{Li}-(np)]) = 
(1.62\pm 0.75)\times 10^{-2}/K^{-}_{stop}$,
where the statistical and systematic errors have been added in quadrature.

The full circles refer to
the semi-inclusive QF $\Sigma^- p$ reaction (F set), while the full triangles 
are inclusive
measurements from the E set. 
The error bars report the statistical uncertainty; the total error corresponds 
to the height of the shaded box (of arbitrary width).
New and old measurements are in good agreement.

For inclusive measurements the systematic errors are large and any
assessment about the existence of a trend as a function
of the mass number is difficult; 
more interesting indications can be obtained by the 
more precise evaluation of the semi-inclusive QF $\Sigma^-p$ reaction. 
A remarkable decrease of the capture rate is evident for $^7$Li
as compared to $^6$Li. This can be motivated by the presence of an
additional  neutron in $^7$Li which enhances the probability
of one-nucleon absorption as compared to the surface interaction of
a $K^-$ with a $(np)$ pair. 
An analogous observation was made in the case of 
$\Sigma^\pm\pi^\mp$ productions 
\cite{re:sigmapi}, which both proceed via one-proton absorption of the
kaon. 
The enhancement of the capture rate for $^9$Be and $^{16}$O 
can be accounted for considering the statistical increase of possible
$(np)$ absorption centers with the nucleus mass number.
The average rates, over the $6 \le A \le 16$ mass number range, are 
$(0.69\pm 0.05)\times 10^{-2}/K^-_{stop}$ for the 
semi-inclusive reactions, and 
$(3.94\pm 0.36)\times 10^{-2}/K^-_{stop}$  
for the inclusive ones, respectively. These new precision measurements
can be fruitfully exploited, after proper extrapolations from 
the present ordinary 
temperature and density, as inputs to improve the current 
microscopic description of the hyperonic-nuclear matter EoS, crucial for
a better understanding of dense astrophysical objects like compact stars
\cite{re:hypstars}.
These new data can also be useful to disclose the onset of excited baryonic
states, like $\Lambda(1405)$ and $\Sigma(1385)$, in dense  nuclear
matter.

In conclusion, for the first time an assessment
of the emission rates of the 
$K^- A\rightarrow \Sigma^- p A^\prime$ reaction on some 
{$p$-shell}
nuclei, with mass number in the range $6 \le A \le 16$, has been performed,
complementing the quite few existing available measurements.
It is shown that a sizeable part of the experimental spectra 
fails to be explained by the 
simple QF two-nucleon $\Sigma^- p$
absorption reaction. Only an accurate spectral analysis
would be able to disentangle the different contributions to the whole 
phase space volume, fully accounting for both contaminating reactions
and other kaon induced reactions with $\Sigma^- p$ production \cite{re:exa14}.
$\Sigma^- p$ additional pairs can in fact be produced together 
with other undetected particles (pions and
nucleons), and they may also come from the decay of more 
massive baryonic resonances. 

Unfortunately, no experiment exists to-date or is 
planned which will provide data of kaon 
induced reactions on the mentioned $p$-shell nuclei, 
comparable to the FINUDA data sample and to the same level of accuracy. 
The presented data are so far unique to 
get insight on the low energy
kaon interactions in nuclei and, in general, in the strangeness nuclear physics
scenario.

\bigskip\bigskip

\acknowledgments{The Authors would like to thank A. Gal and 
J. Mare$\check{\mathrm{s}}$ for the careful
reading of the manuscript and fruitful comments, 
suggestions and discussions.}


\begin{thebibliography}{199}
\bibitem{re:katz} P.A. Katz {\it et al.}, {\it Phys. Rev.} {\bf D1} (1970), 1267 
\bibitem{re:e549} T. Suzuki {\it et al.}, {\it Phys. Rev.} {\bf C76} (2007), 068202
\bibitem{re:emulsio}
W.L. Knight {\it et al.}, {\it Nuovo Cim.} {\bf 32} (1964), 598 \\
H. Davis {\it et al.}, {\it Nuovo Cim.} {\bf 53A} (1968), 313 
\bibitem{re:bubble}
C. Vander Velde-Wilquet {\it et al.}, {\it Nuovo Cim.} {\bf 39} (1977), 538 \\
J.W. Moulder {\it et al.}, {\it Nucl. Phys.} {\bf 35B} (1971), 332 
\bibitem{re:germanoplb}
FINUDA Collaboration, M. Agnello {\it et al.}, {\it Phys. Lett.} {\bf B698} (2011), 219
\bibitem{re:HypX}  FINUDA Collaboration, M. Agnello {\it et al.}, {\it Nucl. Phys.} {\bf A835} (2010), 439
\bibitem{re:hypstars} D. Lonardoni, A. Lovato, S. Gandolfi, F. Pederiva, {\it Phys. Rev. Lett.} {\bf 114} (2015), 092301 \\
D. Lonardoni, F. Pederiva, S. Gandolfi {\it Phys. Rev.} {\bf C89} (2014), 014314\\
D. Lonardoni, S. Gandolfi, F. Pederiva, {\it Phys. Rev.} {\bf C87} (2013), 041303(R) \\
H.-J. Schulze, T. Rijken, {\it Phys. Rev.} {\bf C84} (2011), 035801\\
I. Vida\~na {\it et al.}, {\it Eur. Phys. Lett.} {\bf 94} (2011), 11002 \\
I. Vida\~na, I. Bombaci, A. Polls, A. Ramos, {\it Astr. Astr.} {\bf 339} (2002), 687 
\bibitem{re:yama}
Y. Akaishi, T. Yamazaki, {\it Phys. Rev.} {\bf C65} (2002), 044005 \\
T. Yamazaki, Y. Akaishi, {\it Nucl. Phys.} {\bf B535} (2002), 70
\bibitem{re:finudaprl} M. Agnello {\it et al.}, {\it Phys. Rev. Lett.} {\bf 94} (2005), 212303
\bibitem{re:ox} 
G. Bendiscioli {\it et al.}, {\it Eur. Phys. J.} {\bf A40} (2009), 11 \\
G. Bendiscioli {\it et al.}, {\it Nucl. Phys.} {\bf A789} (2007), 222
\bibitem{re:disto}
DISTO Collaboration, T. Yamazaki {\it at al.}, {\it Phys. Rev. Lett.} {\bf 104} (2010), 132502
\bibitem{re:fopi}
FOPI Collaboration, K. Suzuki {\it et al.}, {\it Prog. Theor. Phys. Suppl.}
{\bf 186} (2010), 351 
\bibitem{re:FINUDAnew} 
FINUDA Collaboration, M. Agnello {\it at al.}, {\it Nucl. Phys.} {\bf A914} (2013), 310
\bibitem{re:oset}
W. Weise, H. H\"artle, {\it Nucl. Phys.} {\bf B535} (2002), 70 \\
N. Shevchenko, A. Gal, J. Mare$\check{\mathrm{s}}$, {\it Phys. Rev. Lett.} 
{\bf 98} (2007), 082301 \\
A. Ramos, E. Oset, {\it Nucl. Phys.} {\bf B671} (2000), 153
\bibitem{re:sigmapi}
FINUDA Collaboration, M. Agnello {\it et al.}, {\it Phys. Lett.} {\bf B704} (2011), 474
\bibitem{re:heusi} 
B. Loiseau, S. Wycech, {\it Phys. Rev.} {\bf C63} (2001), 034003 \\
P. Heusi {\it et al.}, {\it Nucl. Phys.} {\bf A407} (1983), 429
\bibitem{re:fermi} E.J. Moniz {\it et al}, {\it Phys. Rev. Lett.} {\bf 26} (1971), 445
\bibitem{re:finuda1} FINUDA Collaboration, M. Agnello {\it et al.}, {\it Nucl. Phys.} {\bf A775} (2006), 35
\bibitem{re:apparato} 
FINUDA Collaboration, M. Agnello {\it et al.}, {\it Phys. Lett.} {\bf B622} (2005), 53 
\bibitem{re:li6pioni} 
J. Favier {\it et al.}, {\it Nucl. Phys.} {\bf A169} (1971), 540 \\
C. Cernigoi {\it et al.}, {\it Nucl. Phys.} {\bf A352} (1981), 343;
{\it Nucl. Phys.} {\bf A456} (1986), 599
\bibitem{re:yama6li} Y. Akaishi, T. Yamazaki, {\it Nucl. Phys} {\bf A792} (2007), 229
\bibitem{re:tofino} V. Filippini, M. Marchesotti, C. Marciano, {\it Nucl. Instr. Meth.} {\bf A424} (1999), 343
\bibitem{re:isim} P. Bottan {\it et al.}, {\it Nucl. Instr. Meth.} {\bf A427} (1999), 423
\bibitem{re:lmd} 
M. Agnello {\it et al.}, {\it Nucl. Instr. Meth.} {\bf A385} (1997), 58 \\
M. Agnello {\it et al.}, {\it Nucl. Instr. Meth.} {\bf A452} (2000), 386
\bibitem{re:straw}  L. Benussi {\it et al.}, {\it Nucl. Instr. Meth.} {\bf A361} (1995), 180; \\
L. Benussi {\it et al.}, {\it Nucl. Instr. Meth.} {\bf A419} (1998), 648
\bibitem{re:LambdaD}
FINUDA Collaboration, M. Agnello {\it et al.}, {\it Phys. Lett.} {\bf B654} (2007), 80
\bibitem{re:LambdaT}
FINUDA Collaboration, M. Agnello {\it et al.}, {\it Phys. Lett.} {\bf B669} (2008), 229
\bibitem{re:tofone} A. Pantaleo {\it et al.}, {\it Nucl. Instr. Meth.} {\bf A545} (2005), 593
\bibitem{re:tesiBarbara} B. Dalena, Ph. D. Thesis, Universit\`a degli Studi
di Bari, 2006 \hfill\break 
(http://inspirehep.net/record/712914/files/712914.pdf)
\bibitem{re:geane} V. Innocente,  M. Maire, E. Nagy, 
{\it GEANE: Average Tracking and Error Propagation Package CERN Program Library W5013-E} (1991)
\bibitem{re:nmneutrons}
FINUDA Collaboration, M. Agnello {\it et al.}, {\it Phys. Lett.} {\bf B701} (2011), 556 
\bibitem{re:exa14}
A. Filippi, S. Piano, in {\it Proceedings of the EXA14 Conference}, Vienna, 
Austria (2014), {\it Hyp. 
Interactions}, http://dx.doi.org/10.1007/s10751-015-1174-4, Springer International Publishing (2015)


\end{thebibliography}
\end{document}